\documentclass[manuscript]{aastex63}
\usepackage{graphicx}
\usepackage{subfigure}
\usepackage{amsmath,bm}
\usepackage{lineno}

\begin{document}

\title{Coupling of dynamical tide and orbital motion}
\author{Xing Wei}
\affiliation{IFAA, School of Physics and Astronomy, Beijing Normal University, Beijing 100875, China}
\email{xingwei@bnu.edu.cn}

\begin{abstract}
Dynamical tide consists of various waves that can resonate with orbital motion. We test this coupling of dynamical tide and orbital motion using a simple two-dimensional shallow water model, which can be applied to a rocky planet covered with thin ocean or atmosphere. Then we take the earth-moon system as a fiducial model to calculate the tidal resonances and orbital evolution. We find that tidal dissipation can even increase with increasing orbital separation because of the coupling of dynamical tide and orbital motion. We draw the conclusion that the coupling is not negligible to study the orbital evolution on secular timescale.
\end{abstract}

\section{Motivation}\label{sec:motivation}
The tidal force induces not only the equilibrium tide but also the dynamical tide, e.g., \citet{zahn1975, goldreich1989, savonije2002, ogilvie2004, wu2005a, goodman2009, lai2012}, etc. The former is a large-scale and slow deformation under the hydrostatic balance (i.e., pressure due to deformation is comparable to tidal potential), while the latter consists of various waves, e.g., surface gravity wave ($f$ mode) due to gravity with a free surface, internal gravity wave ($g$ mode) due to buoyancy force with stratification, inertial wave ($r$ mode) due to Coriolis force with rotation, etc. Since dynamical tide is essentially waves, it can resonate with orbital motion, i.e., when orbital frequency is close to one of eigenfrequencies of dynamical tide the tidal response and hence dissipation will greatly increase. When the resonances occur, orbital evolution will speed up because the torque on orbit is proportional to tidal dissipation. For example, the geological evidence from marine sediments and fossils suggests that over the past 0.6 Gyr the earth’s length of day has been increasing at a much greater rate than over the epoch from 2.5 to 0.6 Gyr \citep{Denis2002}, and this might be caused by this coupling.

To test this coupling effect, we will use a simple fluid model, i.e., a two-dimensional shallow water model, which can be applied to thin ocean or atmosphere on a rocky planet surface. In this model, the tide is a mixed mode of surface gravity wave and inertial wave. We will simultaneously solve the fluid equations and orbital evolution equations to take into account the coupling effect. In Section \ref{sec:methods} we give the mathematical model, in Section \ref{sec:results} we show the results, and in Section \ref{sec:summary} a brief summary and discussions are given.

\section{Mathematical model}\label{sec:methods}
We denote the primary by subscript ``1'', the secondary by ``2'', and the orbit by ``o''. The secondary is treated as a mass point so that its angular momentum and spin energy are neglected. The orbital angular momentum $L_o$ and primary's spin angular momentum $L_1$ can be exchanged via tidal torque but the total angular momentum $L=L_o+L_1$ conserves. The total energy $E=E_o+E_1$ decreases via tidal dissipation. We consider an elliptical orbit. The orbital angular momentum is $L_o=M_2[GM_1a(1-e^2)]^{1/2}$ where $a$ is semi-major axis and $e$ eccentricity. The semi-major axis and the orbital frequency $\omega_o$ are related by Kepler's third law $a=(GM/\omega_o^2)^{1/3}$ where $M=M_1+M_2$ is the total mass. Primary's angular momentum is $L_1=I_1\omega_1$ where $I_1$ is moment of inertia and $\omega_1$ spin frequency. The orbital energy is $E_o=-GM_1M_2/(2a)$ and primary's spin energy $E_1=I_1\omega_1^2/2$. The system evolves via the tidal torque $\Gamma$ and the tidal dissipation $D$, which are related through $D=-\Gamma(\omega_o-\omega_1)$. The  equations for the evolution of orbit and primary's spin is
\begin{equation}\label{dynamics}
\dot L_o=\Gamma, \hspace{3mm} \dot L_1=-\Gamma, \hspace{3mm} \dot E_o=\omega_o\Gamma, \hspace{3mm} \dot E_1=-\omega_1\Gamma.
\end{equation}
Sum of the former two equations yields the conservation of total angular momentum $\dot L=0$, and sum of the latter two the dissipation of total energy $\dot E=-D$. We take the earth-moon system for an example to estimate tidal dissipation $D$. The masses are $M_1\approx 5.97\times 10^{27}~{\rm g}$ and $M_2\approx 7.35\times 10^{25}~{\rm g}$. Currently, the orbital parameters are $a\approx 3.83\times 10^{10}~{\rm cm}$ corresponding to $\omega_o=(GM/a^3)^{1/2}\approx 2.68\times 10^{-6}~{\rm s^{-1}}$ and $e\approx 0$, and spin frequency is $\omega_1\approx 7.27\times 10^{-5}~{\rm s^{-1}}$. Using Lunar Laser Ranging (LLR) measurements we find $\dot a\approx 3.8~{\rm cm/yr}$. The rate of orbital energy is estimated as $\dot E_o=-(\dot a/a)E_o=GM_1M_2\dot a/(2a^2)\approx 1.2\times 10^{18}~{\rm erg/s}$ and the tidal torque is $\Gamma=\dot E_o/\omega_o\approx 4.5\times 10^{23}~{\rm erg}$. Thus, the power lost by earth is estimated to be $\dot E_1=-\omega_1\Gamma\approx-3.2\times10^{19}~{\rm erg/s}$, much stronger than $\dot E_o$. Consequently we obtain $D=-(\dot E_o+\dot E_1)\approx3.1\times10^{19}~{\rm erg/s}$.

Equations \eqref{dynamics} can be reduced to
\begin{equation}\label{dynamics1}
\frac{dL_o}{dt}=-\frac{\langle D \rangle}{\omega_o-\omega_1}, \hspace{3mm} \frac{dL_1}{dt}=\frac{\langle D \rangle}{\omega_o-\omega_1}, \hspace{3mm} \frac{d(E_o+E_1)}{dt}=-\langle D \rangle.
\end{equation}
It should be noted that we use the orbital-average $\langle D \rangle$. An elliptical orbit is expressed as $r=a(1-e^2)/(1+e\cos\theta)$ where $r$ and $\theta$ are polar coordinates. Tidal potential $V$ is proportional to $r^{-3}$ such that tidal dissipation $D\propto r^{-6}$ can be written as $D=D_0(r/a)^{-6}$ where $D_0$ is tidal dissipation at $r=a$ of a circular orbit with $e=0$. Inserting the expression of an elliptical orbit into the expression of $D$, we are led to $D=D_0(1-e^2)^{-6}/(1+e\cos\theta)^{-6}$. To obtain the orbit-average $\langle D\rangle$ we integrate $D$ over one orbit, $\langle D\rangle=(1/2\pi)\int_0^{2\pi}Dd\theta$. After a little algebra, we obtain
\begin{equation}\label{averaged-D}
\langle D\rangle=\frac{D_0}{(1-e^2)^6}\left(1+\frac{15}{2}e^2+\frac{45}{8}e^4+\frac{5}{16}e^6\right).
\end{equation}
The numerical procedure is as follows. Given $a$, $e$ and $\omega_1$ at each time step, we calculate $\langle D \rangle$, integrate $dL_1/dt$ to obtain $\omega_1$ at the next time step; integrate $d(E_o+E_1)/dt$ to obtain $a$ at the next time step; and integrate $dL_o/dt$ to obtain $e$ at the next time step. Therefore, by integrating \eqref{dynamics1} we obtain the evolution of $a$, $e$ and $\omega_1$.

As mentioned in Section 1, during orbital evolution dynamical tide can resonate with orbital motion such that tidal dissipation can greatly increase. We choose the shallow water model to study dynamical tide. The shallow water model is usually applied to ocean or atmosphere. In this model the fluid depth is much less than the horizontal lengthscale, and according to the divergence-free condition of fluid velocity we can deduce that radial velocity is much less than horizontal velocity \citep{pedlosky}. The dynamical tide is then governed by Laplace's tidal equations in spherical polar coordinates $(r,\theta,\phi)$ with the frictional terms \citep{wei2019,wei2021}
\begin{equation}\label{tidal}
\left\{
\begin{aligned}
&\frac{\partial u_\theta}{\partial t}-2\omega_1\cos\theta u_\phi=\frac{1}{R_1}\frac{\partial}{\partial\theta}(-g\eta+V+\Psi)-\alpha u_\theta \\
&\frac{\partial u_\phi}{\partial t}+2\omega_1\cos\theta u_\theta=\frac{1}{R_1\sin\theta}\frac{\partial}{\partial\phi}(-g\eta+V+\Psi) -\alpha u_\phi \\
&\frac{\partial\eta}{\partial t}+\frac{h}{R_1\sin\theta}\left(\frac{\partial}{\partial\theta}(u_\theta\sin\theta)+\frac{\partial u_\phi}{\partial\phi}\right)=0
\end{aligned}
\right.	
\end{equation}
where $u_\theta$ and $u_\phi$ are the horizontal components of tidal velocity and $\eta$ is tidal height. The first two equations are about momentum conservation and the last equation is about mass conservation. In the momentum equations, the second terms on LHS are Coriolis force, and the last terms on RHS are the frictional force which is simply assumed to be proportional to velocity, i.e., the so-called Rayleigh drag. The last equation is derived from the mass conservation of incompressible fluid $\bm\nabla\cdot\bm u=0$ together with the free-surface condition $d\eta/dt=u_r$ and the shallow water assumption that horizontal velocity is independent of radius $\partial u_\theta/\partial r=\partial u_\phi/\partial r=0$. The other parameters are as follows: $h$ is primary's ocean or atmosphere depth, $V=(3/4)(M_2/M_1)(R_1/a)^3gR_1\sin^2\theta\cos(2\phi-2\omega t)$ is the time-periodic tidal potential where $\omega=\omega_o-\omega_1$ is tidal frequency, and $\Psi=(3/5)(\rho/\bar\rho)g\eta$ is the self-gravity potential where $\rho$ is fluid density and $\bar\rho$ primary's mean density. At low $\omega_1$, fluid equations \eqref{tidal} were analytically solved in the asymptotic limit $\omega_1^2R_1/g\ll 1$ by \citet{Longuet-Higgins1968}. But during the orbital evolution, $\omega_1$ evolves and fluid equations \eqref{tidal} need be numerically solved by spectral method and the details can be found in \citet{wei2019}.

Now dynamical tide and orbital motion are coupled by tidal dissipation
\begin{equation}\label{D}
D_0=\int_\Omega\rho\alpha(u_\theta^2+u_\phi^2)d\Omega
\end{equation}
where $\Omega$ is the fluid volume. It should be noted that we neglect the contribution of radial velocity $u_r^2$ to tidal dissipation because, as mentioned in the last paragraph about the shallow water model, radial velocity is much less than horizontal velocity. Clearly, tidal dissipation is no longer constant but changes with time, and we need to solve \eqref{dynamics1} and \eqref{tidal} simultaneously. At each time step with given $a$, $e$ and $\omega_1$, we solve \eqref{tidal} to obtain $D_0$ by \eqref{D}; next we obtain $\langle D\rangle$ by \eqref{averaged-D}; and finally we solve \eqref{dynamics1} to obtain $a$, $e$ and $\omega_1$ at the next time step.

\section{Results}\label{sec:results}
\begin{figure}
\centering
\includegraphics[scale=0.4]{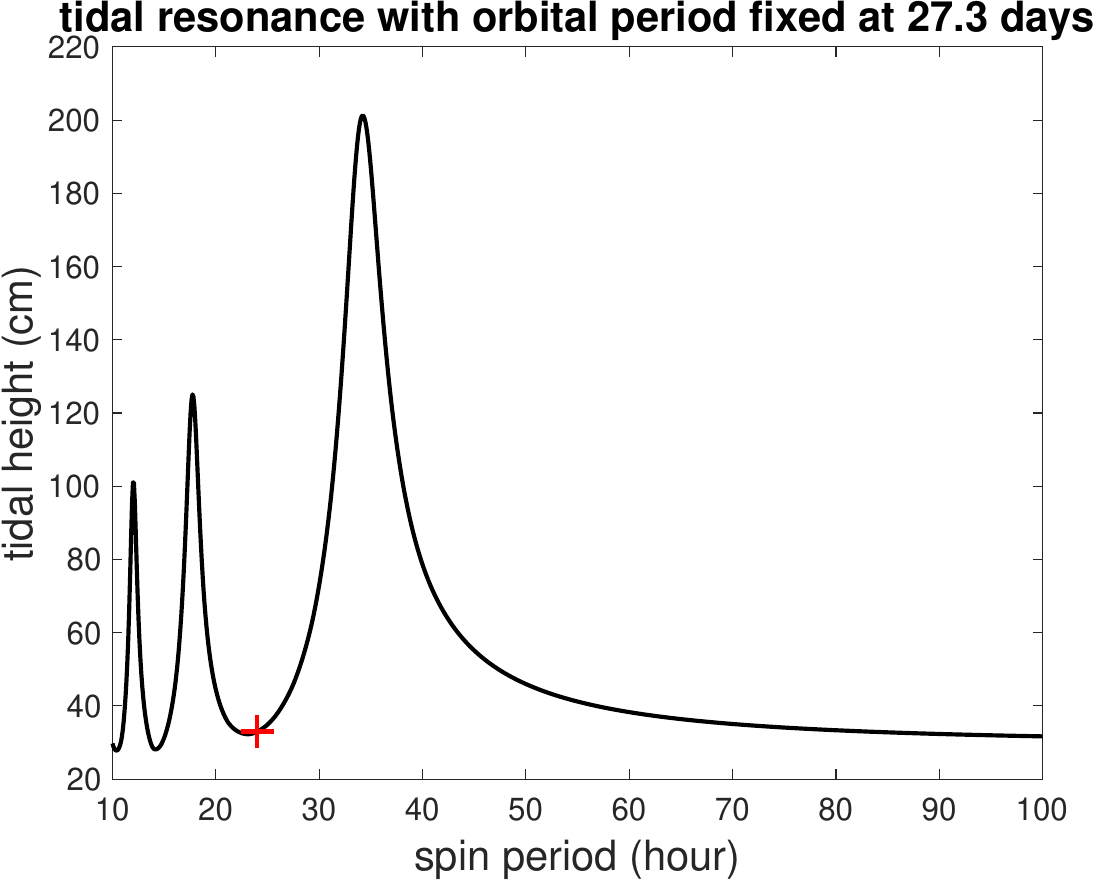}
\includegraphics[scale=0.4]{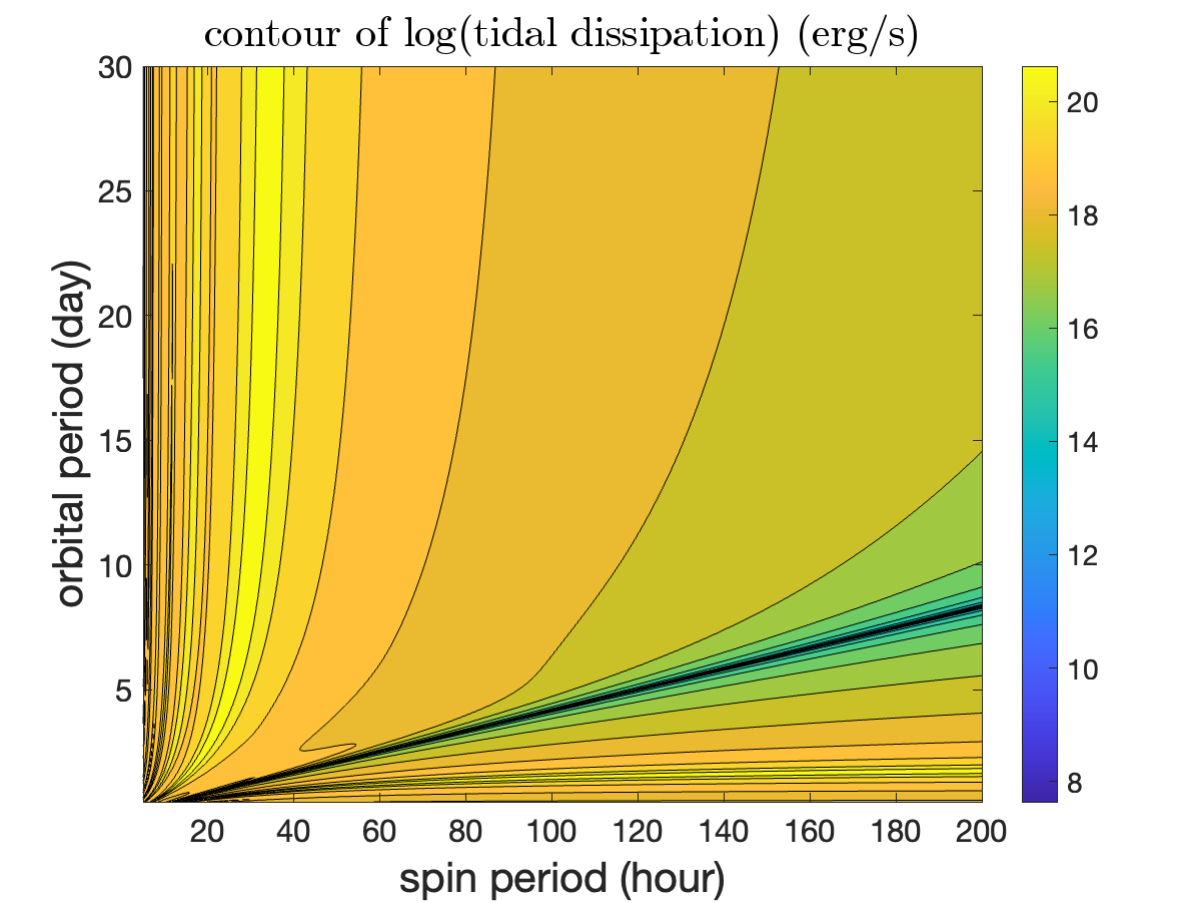}
\caption{Tidal resonance. Left panel: tidal height versus spin period with orbital period fixed (red cross denotes the tidal height nowadays). Right panel: tidal dissipation versus spin period and orbital period.}\label{tidal-resonance}
\end{figure}

In the first place, we study tidal resonances by solely solving fluid equations \eqref{tidal}. We choose the current earth-moon orbit as a fiducial model: $a=3.8\times10^{10}~{\rm cm}$, $e=0.054$, the average ocean depth $h=3688~{\rm m}$, and $\alpha=5.4\times10^{-6}~{\rm s^{-1}}$ to match $\langle D\rangle=3.1\times10^{19}~{\rm erg/s}$. We fix the orbital frequency $\omega_o=2\pi/{\rm (27.3~days)}$ to calculate the tidal height $\eta$ as a function of the spin period $2\pi/\omega_1$. We can see the three peaks of tidal resonances as shown in the left panel of Figure \ref{tidal-resonance}. We are now at the spin period 24 hours which corresponds to the tidal height 33 cm, in agreement of the observation of the global ocean tide far away from continents. Next we calculate the tidal dissipation $D$ as a function of both the spin period $2\pi/\omega_1$ and the orbital period $2\pi/\omega_o$, as shown in the right panel of Figure \ref{tidal-resonance}. The bright yellow area indicates that the strongest resonance occurs at the spin period around 30-40 hours and the orbital period longer than 5 days.

\begin{figure}
\centering
\includegraphics[scale=0.3]{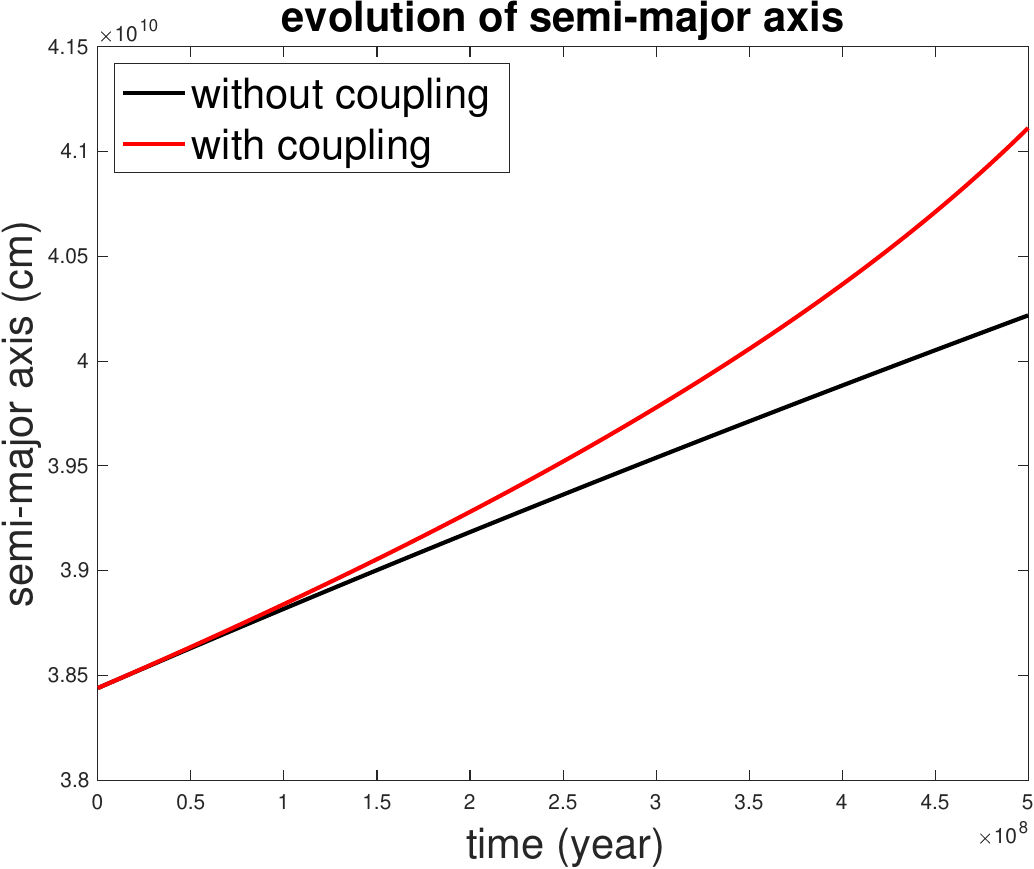}
\includegraphics[scale=0.3]{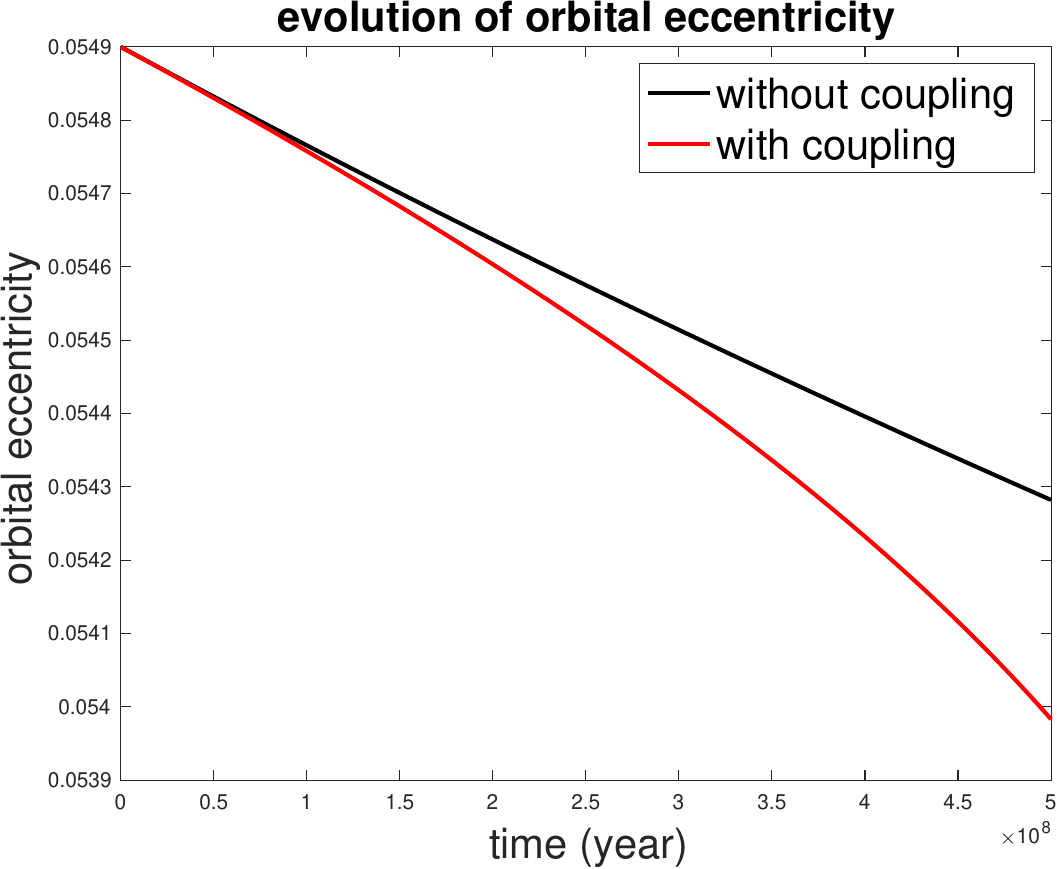}
\includegraphics[scale=0.3]{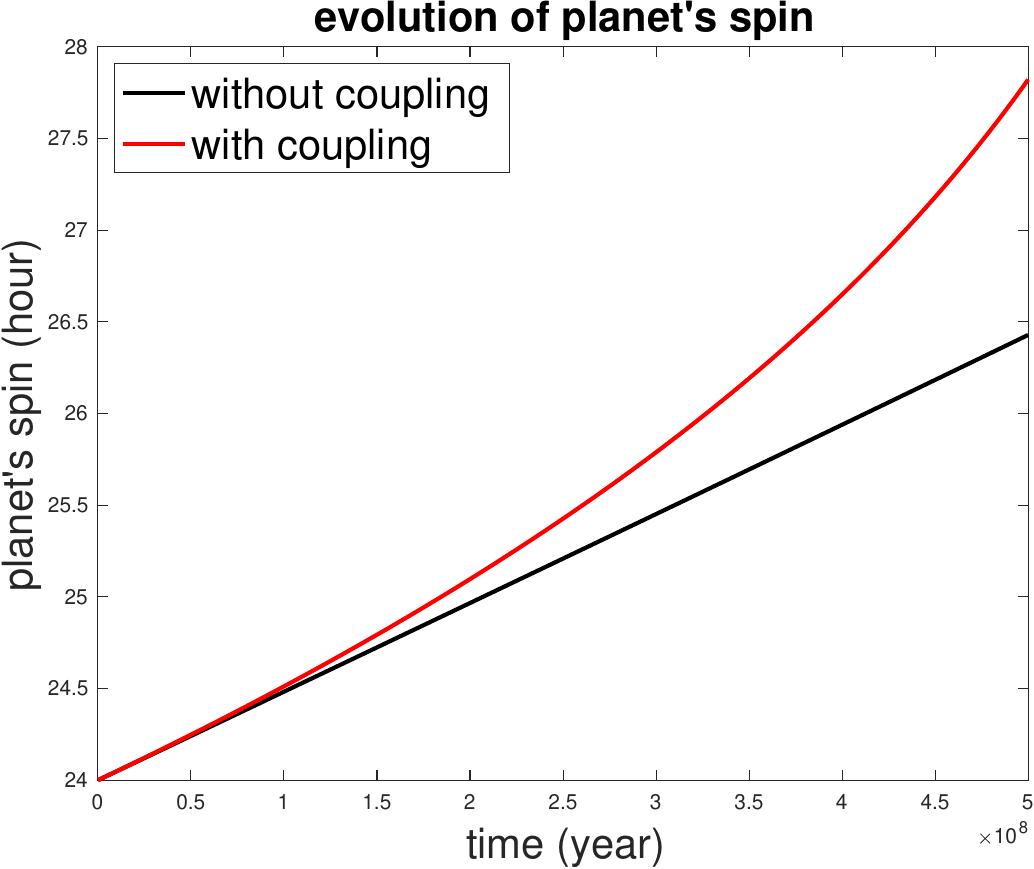}
\caption{Orbital and spin evolution to investigate the future earth-moon. The evolution track of $a$, $e$ and $\omega_1$. Initial $a=3.8\times10^{10}~{\rm cm}$, $e=0.054$ and $\omega_1=2\pi/(24~{\rm hours})$. Black curves without coupling and red with coupling.}\label{orbital-evolution}
\end{figure}

Then we study the coupling effect of dynamical tide and orbital motion by simultaneously solving \eqref{dynamics1} and \eqref{tidal} that are coupled through tidal dissipation. We set the initial $a=3.8\times10^{10}~{\rm cm}$, $e=0.054$ and $\omega_1=2\pi/(24~{\rm hours})$ to calculate the evolution track of $a$, $e$ and $\omega_1$ in the future 0.5 Gyr. Figure \ref{orbital-evolution} shows the results: black curves without coupling and red curves with coupling. Without the coupling, we solve \eqref{dynamics1} with a given $\langle D\rangle$ of its initial value, and it can seen that the orbit and spin ($a$, $e$ and $\omega_1$) evolve with time almost linearly. But with the coupling, we solve \eqref{dynamics1} and \eqref{tidal} simultaneously, and it can be seen that the orbit and spin evolve with time much faster than without the coupling. The results suggest that moon's orbit and earth's spindown will evolve much faster than previously expected without the coupling of dynamical tide and orbital motion. 

\begin{figure}
\centering
\includegraphics[scale=0.3]{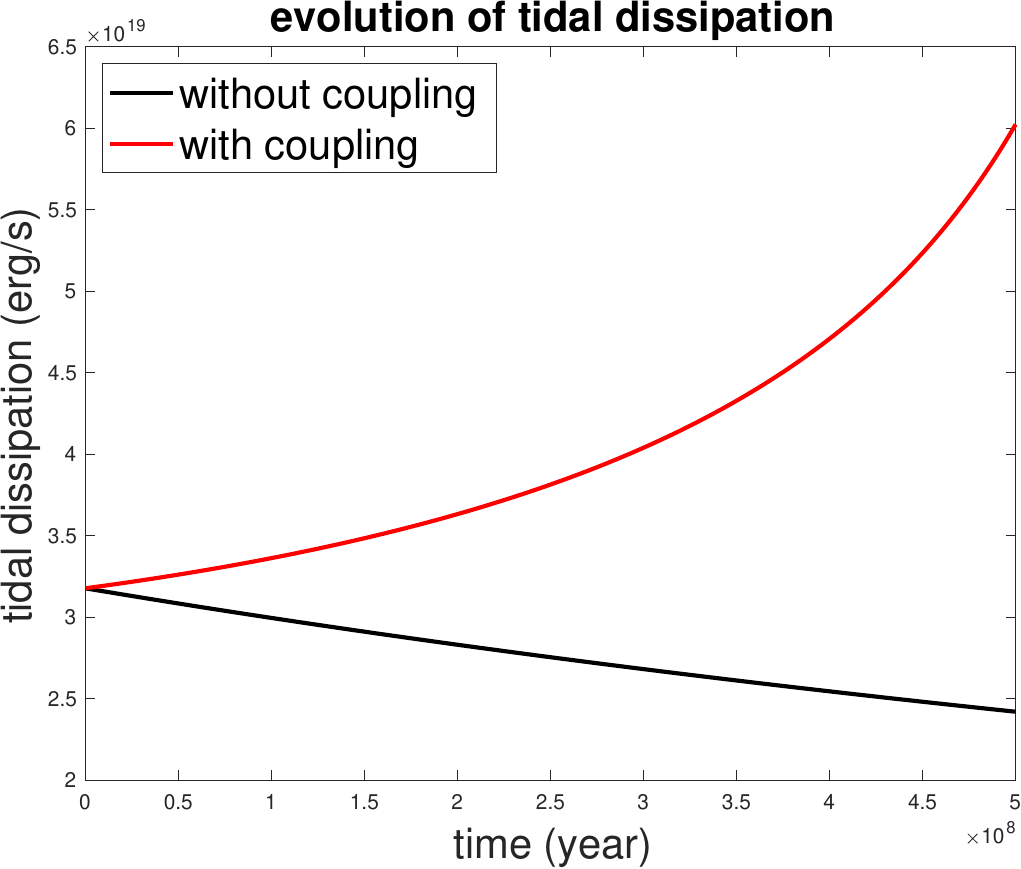}
\includegraphics[scale=0.3]{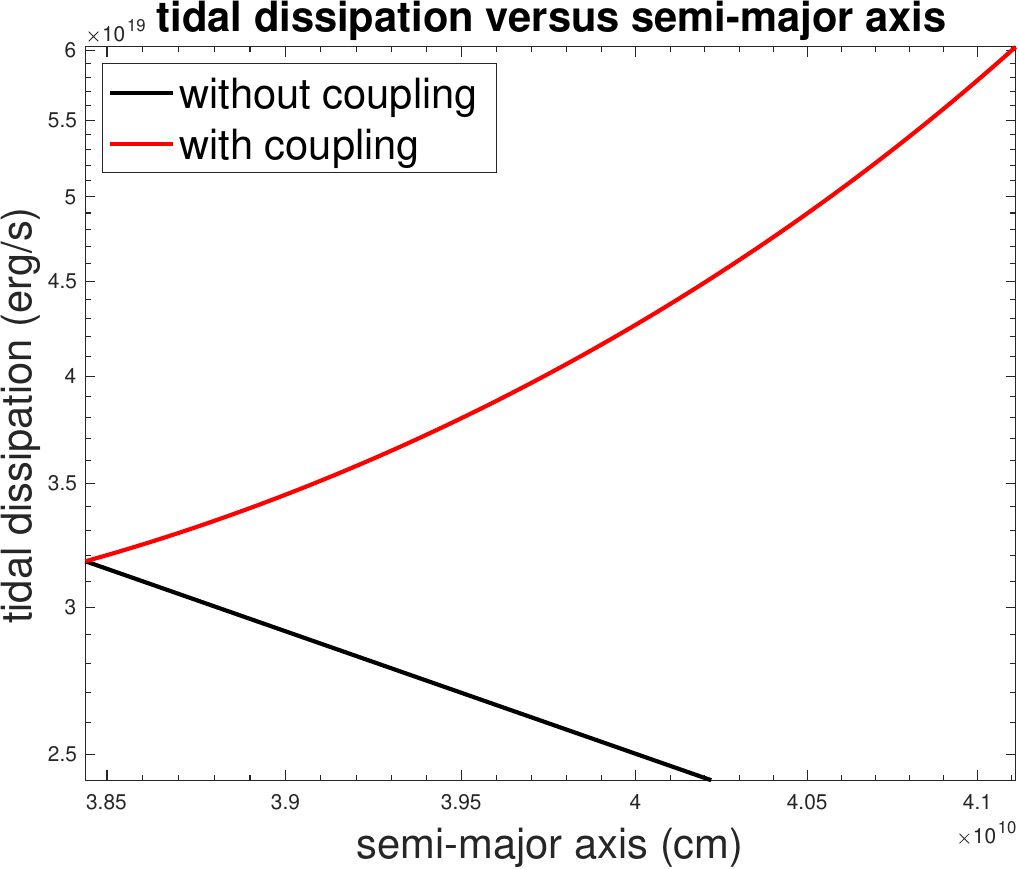}
\caption{Left panel: tidal dissipation versus time. Right panel: dissipation versus semi-major axis. Initial $a=3.8\times10^{10}~{\rm cm}$, $e=0.054$ and $\omega_1=2\pi/(24~{\rm hours})$. Black curves without coupling and red with coupling.}\label{dissipation-evolution}
\end{figure}

Figure \ref{dissipation-evolution} shows the reason of evolution speedup. The left panel shows the evolution of tidal dissipation. It can seen that dissipation decreases with time when the coupling neglected but increases with time when the coupling taken into account. The right panel shows dissipation versus semi-major axis. As we know $\langle D\rangle\propto a^{-6}$, and the black curve without coupling obeys this scaling law. However, the red curve with coupling shows that tidal dissipation increases with semi-major axis $a$. This behaviour definitely arises from the resonances of dynamical tide and orbital motion.

\section{Summary and discussions}\label{sec:summary}
In this short paper we build a model, i.e., the orbital equations coupled with the fluid equations through tidal dissipation, to study the coupling of dynamical tide and orbital motion. We find that tidal dissipation can increase with increasing orbital semi-major axis because of this coupling effect. On a secular timescale, this coupling effect should be taken into account. In our model, the orbital inclination is not considered, and if inclination is taken into account then orbital equations \eqref{dynamics1} should be replaced by \eqref{dynamics}. We use a 2D fluid model and readers can extend to a more complicated 3D fluid model for the other cases, e.g., gaseous planets. Readers can extend our model to exoplanet-exomoon systems when more observational data are available.

\bibliography{paper}{}
\bibliographystyle{aasjournal}

\end{document}